# Universality in the level statistics of disordered systems

Heiko Bauke[*] and Stephan Mertens[†]
*Institut für Theoretische Physik, Otto-von-Guericke Universität,
PF 4120, 39016 Magdeburg, Germany*

Energy spectra of disordered systems share a common feature: if the entropy of the quenched disorder is larger than the entropy of the dynamical variables, the spectrum is locally that of a random energy model and the correlation between energy and configuration is lost. We demonstrate this effect for the Edwards-Anderson model, but we also discuss its universality.

Energy levels in disordered systems are quenched random variables with distribution induced by the random couplings between the dynamical variables. In general the energy levels are correlated, but under certain conditions these correlations can be neglected. A well known example is the $p$-spin generalization of the Sherrington-Kirkpatrick model [1], where the correlations decrease with increasing $p$ [2, 3]. In the large $p$ limit, the energy levels can be treated as *independent* random variables, the corresponding model is called *random energy model* or REM [2]. In a REM the role of the dynamical variables $\sigma$ is reduced to that of indices in a table of uncorrelated random energy values. In this contribution we will argue that the *local* structure of the energy spectrum in many disordered systems is that of a REM: energy levels that are neighbors on the energy axis behave like uncorrelated random variables, and the corresponding configurations are uncorrelated, too. We will demonstrate this explicitly for the Edwards-Anderson spin glass, but the mechanism behind this local REM is much more general.

The Edwards-Anderson (EA) model [4] is the paradigmatic model in spin glass physics [5]. Its energy is defined as

$$\hat{H}(\sigma) = -\sum_{\langle i,j \rangle} J_{ij} \sigma_i \sigma_j \tag{1}$$

where the $N$ Ising spins $\sigma_i = \pm 1$ are located on a regular lattice, the sum is over all nearest-neighbor pairs of the lattice, $J_{ij}$ denotes the coupling between spins $\sigma_i$ and $\sigma_j$. The $J_{ij}$ are independent, identically distributed (i.i.d.) random variables with second moment $\Delta$. We will assume that the probability density of $J_{ij}$ is a piecewise continuous function on the real axis. This includes the Gaussian as well as the uniform distribution but excludes for example the bimodal distribution. In order to get an expression for the density of states that is asymptotically independent of $N$, $\Delta$ and the dimension of the lattice we introduce a scaling factor $(N\nu\Delta)^{-1/2}$ where $\nu$ is the number of bonds per spin, i.e. we consider

$$H(\sigma) = -\frac{1}{\sqrt{N\nu\Delta}} \sum_{\langle i,j \rangle} J_{ij} \sigma_i \sigma_j \tag{2}$$

instead of $\hat{H}$. The prefactor ensures that the asymptotic density of states is a simple Gaussian,

$$g(E) = \frac{1}{2^N} \sum_\sigma \delta(E - H(\sigma)) \simeq \frac{1}{\sqrt{2\pi}} e^{-E^2/2}. \tag{3}$$

For finite $N$ we can number the levels in ascending order,

$$\ldots < E_{-1} < E_0 < 0 < E_1 < E_2 < \ldots. \tag{4}$$

Let us consider all levels inside a small reference interval $[\alpha, \alpha + d\alpha]$ for fixed $\alpha$. Let the interval contain $M$ levels $E_{r+1}, E_{r+2}, \ldots, E_{r+M}$ where $r = r(\alpha)$ is defined by

$$E_r < \alpha \leq E_{r+1}. \tag{5}$$

Now let us *assume* that the $M$ levels inside our interval are statistically independent. Since $M = \mathcal{O}(2^N)$ is large we can apply asymptotic order statistics [6] to see that for any fixed $\ell \geq 1$ the scaled tuple

$$\frac{2^{N-1}}{\sqrt{2\pi}} e^{-\alpha^2/2} \left[(E_{r+1}, E_{r+2}, \ldots, E_{r+\ell}) - \alpha\right] \tag{6}$$

converges in distribution to $(W_1, W_1 + W_2, \ldots, W_1 + \cdots + W_\ell)$, where $W_i$ are i.i.d. random variables with exponential distribution, $p(W_i) = e^{-W_i}$. From this one can derive the distributions $p_k(\varepsilon_k)$ of the scaled energies

$$\varepsilon_k(\alpha) = \lim_{N \to \infty} \frac{2^{N-1}}{\sqrt{2\pi}} e^{-\alpha^2/2} (E_{r+k} - \alpha) \tag{7}$$

$k = 1, \ldots, \ell$ to be

$$p_k(\varepsilon_k) = \frac{\varepsilon_k^{k-1}}{\Gamma(k)} e^{-\varepsilon_k}. \tag{8}$$

To check our assumption of statistically independent energy levels we measured the distribution of $\varepsilon_k$ by exhaustive enumerations of the EA-model on finite square lattices. The data confirm eq. (8) even for moderate values of $N$, as can be seen in fig. 1.

Apparently the energy values in an interval $[\alpha, \alpha + d\alpha]$ are asymptotically uncorrelated. To check whether this is true for the corresponding *configurations* as well we consider the overlap

$$q(\sigma, \sigma') = \frac{1}{N} \left| \sum_{j=1}^N \sigma_j \sigma'_j \right| \tag{9}$$

between two configurations $\sigma$ and $\sigma'$ with energies $E(\sigma) = E_r$ and $E(\sigma') = E_{r+1}$. For an ensemble of random instances we

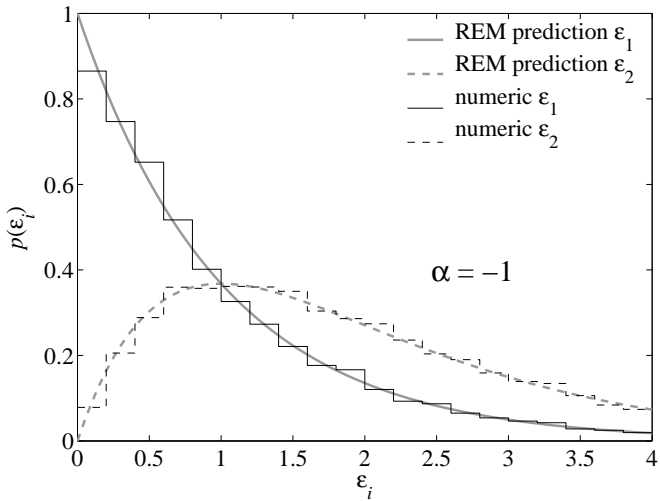

Figure 1: Distribution of $\varepsilon_1$ and $\varepsilon_2$ at $\alpha = -1$ for the EA-model on a $6 \times 5$ square lattice. The numerical distribution was calculated by averaging over 10 000 random realizations of couplings. Data for different values of $\alpha$ look very similar.

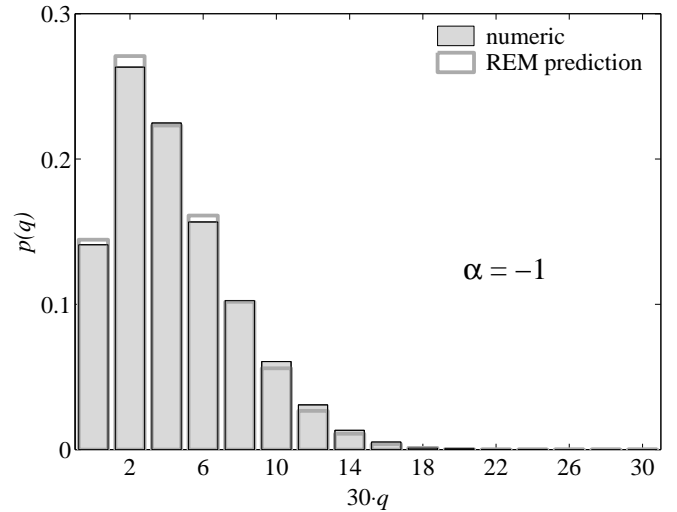

Figure 2: Distribution of the overlap between two configurations with adjacent energy values at $\alpha = -1$ for the EA-model on a $6 \times 5$ square lattice. The numerical distribution was calculated by averaging over 10 000 random realizations of couplings. Data for different values of $\alpha$ look very similar.

get a distribution $p(q)$ that should be given by

$$p(q) = \begin{cases} \frac{1}{2^N} \binom{N}{N(1-q)/2} & \text{for } q = 0, \\ \frac{2}{2^N} \binom{N}{N(1-q)/2} & \text{for } q > 0. \end{cases} \quad (10)$$

if $\sigma$ and $\sigma'$ were uncorrelated. Again this is confirmed by the numerics, see fig. 2. Our numerical experiments corroborate the claim that the energy spectrum of the EA-model is locally that of a REM. We call this the local REM property.

Before we discuss the origin of the local REM property let us mention that it has been established rigorously for the one dimensional case. The Hamiltonian of the random bond Ising chain can be written as

$$H(\tau) = \frac{1}{\sqrt{N\Delta}} \sum_{j=1}^{N} |J_j| \tau_j \quad (11)$$

where we have introduced new Ising variables $\tau_j := -\text{sgn}(J_j)\sigma_j\sigma_{j+1}$. The absolute value $|H(\tau)|$ is the cost function of the number partitioning problem (NPP), a classical problem from combinatorial optimization. The local REM property of the NPP [7] has been rigorously proven [8, Theorem 2.8] for the low energies, i.e. for $\alpha = 0$, but recent numerical studies have confirmed its validity for $\alpha > 0$, too [9].

The origin of the local REM is best understood in terms of the bit-entropy of the couplings $J_{ij}$. Let us consider the case where each $J_{ij}$ is a random integer, uniformly drawn from the set $\{0, 1, \ldots, 2^B - 1\}$. Typical energies of the Hamiltonian

$$\hat{H}(\sigma) = -\sum_{\langle i,j \rangle} J_{ij}\sigma_i\sigma_j \quad (12)$$

are integers in the interval $-\sqrt{vN}2^B \ldots \sqrt{vN}2^B$. Now let us assume that $B \ll N$. Then the number of configurations exceeds the number of available energy levels by far. All potential energy levels in the interval $-\sqrt{vN}2^B \ldots \sqrt{vN}2^B$ are populated, i.e. the spacing of adjacent energy levels is deterministically fixed to 2. In addition each of these energy levels is exponentially degenerated. In the NPP this is called the "easy phase" [10] because it is algorithmically relatively easy to find *one* configuration with a predetermined energy $0 \leq E < \sqrt{vN}2^B$. The same holds for spin glass Hamiltonians and $-\sqrt{vN}2^B < E < \sqrt{vN}2^B$. In the other regime with $B \gg N$ the levels are no longer degenerated (except the twofold degeneracy due to the overall spin flip symmetry), and finding a unique configuration with energy that comes as close as possible to a predetermined energy is algorithmically very hard. In the NPP this is called the "hard" phase. With $\mathcal{O}(2^N)$ different energy values we get a spacing of $\mathcal{O}(\sqrt{N}2^{B-N})$ between adjacent levels, but the *precise* value of the spacing within this range is determined by the low significant bits in the $J_{ij}$ and cannot be controlled by flipping spins. Now assume that you have a configuration $\sigma$ with energy $E_r$ and you want to find the configuration $\sigma'$ that brings you to the next larger energy $E_{r+1}$. Flipping a single spin in $\sigma$ changes the energy at least by $\mathcal{O}(2^B/N)$, which is the order of magnitude of the minimum of $|J_{ij}|$. Hence a single spin flip initially brings you far away from the target. The same is true for any finite number of spin flips. Reaching a level at distance $\mathcal{O}(\sqrt{N}2^{B-N})$ requires the concerted adjustment of $\mathcal{O}(N)$ spins. And the spins to be flipped are again determined by the uncontrollable low significant random bits in the $J_{ij}$. To put a long story short we expect the local REM property to hold whenever the bit entropy in the quenched disorder ($B$) exceeds the bit entropy of the dynamical variables ($N$).

Our reasoning is very hand-wavy, but at least it leads to a conclusion that can be checked experimentally: the local REM property should depend on the bit-entropy of the quenched disorder. For the NPP this has been proven rigorously [8]. For the EA-model we did the following experiment: we generated the $J_{ij}$ as random $B$-bit integers, specify a reference energy $\alpha$ and calculate all configurations $\{\sigma\}$ that have $E(\sigma) = E_{r+1}$ and all configurations $\{\sigma'\}$ that have $E(\sigma') = E_{r+2}$. Then we identified the maximum $q_{max}$ of all overlaps $q(\sigma, \sigma')$ of configurations from these two sets. This is repeated for many random instances with varying $B$. According to our heuristic reasoning we expect $q_{max} = 1 - 2/N$ for $B \ll N$ independently of the $J_{ij}$. Due to the large degener-acy there is always a way to go from one energy level to the next by a single spin flip. For $B \gg N$ the sets $\{\sigma\}$ and $\{\sigma'\}$ consists of a single configuration each (neglecting spin-flip symmetry), and the overlap between these configurations does depend on the quenched disorder and should be distributed according to eq. (10). Fig. 3 shows that this is precisely observed in the numerical experiment. The transition between the two regimes happens at values $B \lesssim N$, and the transition seems to get sharp as $N \to \infty$. The situation is very similar to the "easy-hard" transition in the NPP, where the critical ratio $B/N$ is $1 - \mathcal{O}(\log(N)/N)$ [8, 10].

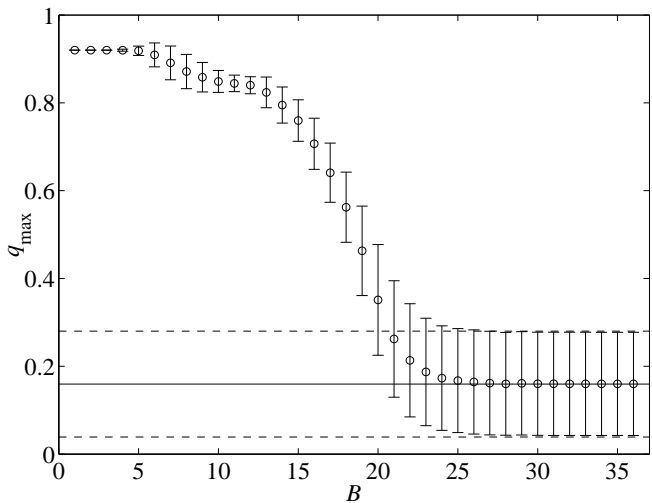

Figure 3: Mean (symbols) and standard deviation (error bars) of the maximum overlap between two configurations with energies $E_1$ and $E_2$ ($\alpha = 0$) for the EA-model on a $5 \times 5$ square lattice. $B$ is the number of bits in the $J_{ij}$, horizontal lines indicate the average (solid) and the standard deviation (dashed) as predicted by the local REM hypothesis, eq. (10). Numerical data are based on random instances for each value of $B$.

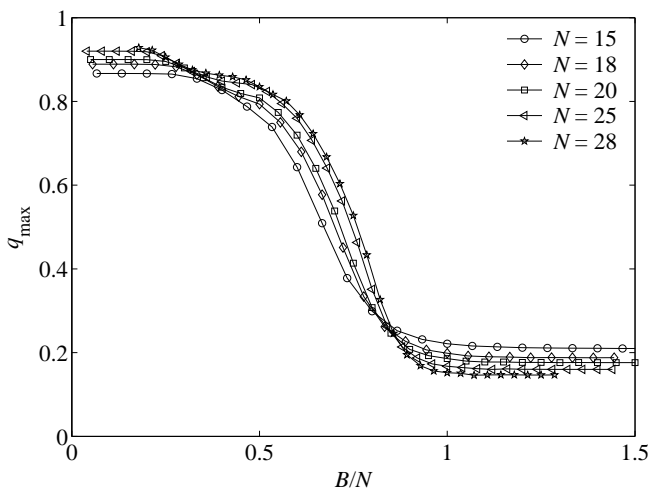

Figure 4: Same experiment as in fig. 3 but for different values of $N$.

This interplay between the "more significant" noise that can be controlled by the dynamical variables and the "less significant noise" that cannot is a rather general phenomenon. The only prerequisite is an energy that is the sum of independent random numbers whose bit-entropy exceeds the number of bits to specify a configuration. We have investigated other spin glass systems like the EA-model in higher dimensions, the Sherrington-Kirkpatrick model, Potts glasses and $p$-spin models[11]. We found that all these systems share the local REM property. The same is true for random instances of classical optimization problems like the Travelling Salesman Problem and the Minimum Spanning Tree Problem. So far we haven't found a disordered system without the local REM property.

Our heuristic explanation suggests that the local REM should hold only for those parts of the spectrum where the interlevel spacing is $\mathcal{O}(2^{-N})$. For many systems this *excludes* energies that are relevant in the low temperature regime. The one dimensional random bond Ising chain (11) is a simple example: ground state and first excitation are highly correlated. They differ by a single spin $\tau_k$ only, $k = \arg\min_j\{|J_j|\}$. The corresponding energy difference is $\mathcal{O}(N^{-3/2})$, which is obviously not dominated by the low order bits of the random couplings. Similar arguments apply to more complicated systems like the SK-model, for which it is known that the overlap distribution in the spin glas phase differs from the overlap distribution of the REM. With its restriction to the paramagnetic phase the local REM is not very interesting in terms of thermodynamic properties of spin glass models. It is more relevant in combinatorial optimization, where it is related to the approximability properties of computationally hard problems. Note that many easy (i.e. polynomial time solvable) optimization problems become NP-hard if one replaces the search for the minimum by the search for the configuration that has an energy as close as possible to a given reference energy $\alpha$. It is an open question how this transition in computational complexity is related to the appearance of the local REM property for certain values of $\alpha$.

The local REM property seems to be universal feature that can be found in all systems with real-valued disorder and in those parts of their energy spectrum with exponentially small level spacing. The underlying mechanism is the dominance of the low order bits in the quenched disorder on the local properties of the energy spectrum. The generality of this mechanism might help to put the local REM hypothesis into a rigorous framework, following the footsteps of [8].

All numerical simulations have been done on our Beowulf cluster TINA, see http://tina.nat.uni-magdeburg.de. This work was supported by Deutsche Forschungsgemeinschaft under grant ME2044/1-1.

* E-mail: heiko.bauke@physik.uni-magdeburg.de
† E-mail: stephan.mertens@physik.uni-magdeburg.de